\documentclass[conference]{IEEEtran}
\IEEEoverridecommandlockouts
\usepackage{cite}
\usepackage{amsmath,amssymb,amsfonts}
\usepackage{algorithmicx}
\usepackage{graphicx}
\usepackage{textcomp} 
\usepackage{xcolor}
\usepackage{lipsum}
\usepackage{xspace}
\usepackage{cleveref}
\usepackage{listings}
\usepackage{xcolor}
\usepackage{comment}
\usepackage{flushend}
\usepackage{subcaption}

\definecolor{codegreen}{rgb}{0,0.6,0}
\definecolor{codegray}{rgb}{0.5,0.5,0.5}
\definecolor{codepurple}{rgb}{0.58,0,0.82}

\lstdefinestyle{CStyle}{
    language=C,
    commentstyle=\color{codegreen},
    keywordstyle=\color{magenta},
    keywordstyle=[1]\color{blue!80!black},    
    keywordstyle=[2]\color{orange!80!black},     
    keywordstyle=[3]\color{red!50!black}, 
    morekeywords={NULL, BLOCK_SIZE},
    morekeywords=[1]{int8_t, uint32_t, int32_t, uint64_t, T, dpu_set, dpu_set_t},
    morekeywords=[2]{ifdef, elif, endif},
    morekeywords=[3]{__builtin_mul_sl_sl_rrr, __builtin_mul_sh_sl_rrr, mram_read, mram_write, barrier_wait, timer_start, update, dpu_alloc_ranks, transfer_data, alloc_buffer, alloc_buffer_on_cpu, timer_stop, equal_channel_distribution, dpu_launch, transfer_output_data},
    numberstyle=\tiny\color{codegray},
    stringstyle=\color{codepurple},
    basicstyle=\ttfamily\footnotesize,
    breakatwhitespace=false,         
    breaklines=true,                 
    captionpos=b,                    
    keepspaces=true,                 
    numbers=left,                    
    numbersep=5pt,                  
    showspaces=false,                
    showstringspaces=false,
    showtabs=false,                  
    tabsize=2,
    breakindent=23em
}

\lstdefinelanguage[RISC-V]{Assembler}{
  alsoletter={.}, 
  alsodigit={0x}, 
  morekeywords=[1]{ 
    move, mul_step, jgtu
  },
  morekeywords=[2]{ 
    __mulsi3_exit, __mulsi3_swap, __mulsi3_start
  },
  morekeywords=[3]{ 
    zero, r0, r1, r2, r3, r4, r5, r6, r18, r19, d0, d1, d2, d3, d4, d5, d6, z, \%2, \%1, true
  },
  morecomment=[l]{;},   
  morecomment=[l]{\#},  
  morestring=[b]",      
  morestring=[b]'       
}

\lstdefinelanguage{UPMEM-OP}{
    morekeywords={op, if, jump},
    sensitive=true,
    morecomment=[l]{\#}, 
    morestring=[b]"",
    morekeywords=[3]{ 
        dc, ra, db, shift, boot_cc, pc, ZF
    },
    directivestyle=\color{red!50!black}\bfseries,
}

\usepackage{algorithm}
\usepackage[noend]{algpseudocode}

\def\BibTeX{{\rm B\kern-.05em{\sc i\kern-.025em b}\kern-.08em
    T\kern-.1667em\lower.7ex\hbox{E}\kern-.125emX}}
\begin{document}

\title{UPMEM Unleashed: Software Secrets for Speed}

\newcommand{\etal}{\emph{et al.}\xspace}
\newcommand{\upmem}{UPMEM\xspace}
\newcommand{\upmemmodel}{UPMEM\mbox{-}v1B\xspace}
\newcommand{\intf}{\texttt{INT4}\xspace}
\newcommand{\uintf}{\texttt{UINT4}\xspace}
\newcommand{\inte}{\texttt{INT8}\xspace}
\newcommand{\uinte}{\texttt{UINT8}\xspace}
\newcommand{\inttt}{\texttt{INT32}\xspace}
\newcommand{\uinttt}{\texttt{UINT32}\xspace}
\newcommand{\intsf}{\texttt{INT64}\xspace}
\newcommand{\float}{\texttt{FP32}\xspace}
\newcommand{\double}{\texttt{FP64}\xspace}
\newcommand{\andinst}{\textsc{AND}\xspace}
\newcommand{\addinst}{\textsc{ADD}\xspace}
\newcommand{\mulinst}{\textsc{MUL}\xspace}
\newcommand{\xor}{\textsc{XOR}\xspace}
\newcommand{\mulslsl}{\texttt{MUL\_SL\_SL}\xspace}
\newcommand{\mulstep}{\texttt{MUL\_STEP}\xspace}
\newcommand{\mulsi}{\texttt{\_\_mulsi3}\xspace}
\newcommand{\leftbitshift}{\textsc{LeftBS}\xspace}
\newcommand{\popcount}{\textsc{PopCount}\xspace}
\newcommand{\addtt}{\texttt{ADD}\xspace}
\newcommand{\multt}
{\texttt{MUL}\xspace}
\newcommand{\inteni}{\texttt{NI}\xspace}
\newcommand{\intenifour}{\texttt{NIx4}\xspace}
\newcommand{\intenieight}{\texttt{NIx8}\xspace}
\newcommand{\intttdim}{\textsc{DIM}\xspace}
\newcommand{\update}{\texttt{update}\xspace}

\newcommand{\todo}[1]{\textbf{TODO: #1}}

\newcommand{\bsdp}{BSDP\xspace}

\author{\IEEEauthorblockN{Krystian Chmielewski,  Jarosław Ławnicki, Uladzislau Lukyanau, Tadeusz Kobus, Maciej Maciejewski}
\IEEEauthorblockA{\textit{Heterogeneous Memory Software Lab} \\
\textit{Huawei Technologies}\\
Warsaw, Poland\\
name.surname@huawei.com}
}

\maketitle

\begin{abstract}
Developing kernels for Processing-In-Memory (PIM) platforms poses unique challenges in data management and parallel programming on limited processing units. Although software development kits (SDKs) for PIM, such as the \upmem SDK, provide essential tools, these emerging platforms still leave significant room for performance optimization. In this paper, we reveal surprising inefficiencies in the \upmem software stack and play with non-standard programming techniques. By making simple modifications to the assembly generated by the \upmem compiler, we achieve speedups of 1.6–2$\times$ in integer addition and 1.4–5.9$\times$ in integer multiplication, depending on the data type. We also demonstrate that bit-serial processing of low precision data is a viable option for  \upmem: in \intf bit-serial dot-product calculation, \upmem can achieve over 2.7$\times$ speedup over the baseline. Minor API extensions for PIM allocation that account for the non-uniform memory access (NUMA) architecture of the server further improve the consistency and throughput of host–PIM data transfers by up to 2.9$\times$. Finally, we show that, when the matrix is preloaded into PIM, our optimized kernels outperform a dual-socket CPU server by over 3$\times$ for \inte generalized matrix-vector multiplication (GEMV) and by 10$\times$ for \intf GEMV. Our optimized \inte GEMV kernel outperforms the baseline 3.5$\times$.

\end{abstract}


\section{Introduction}

The growing disparity between computational throughput and memory bandwidth—often termed the memory wall~\cite{GYK+24}—has intensified interest in Processing-in-Memory (PIM) architectures (see Mutlu \etal \cite{MGGA22} for a survey). By moving computation closer to data, PIM mitigates data movement bottlenecks, promising significant performance and efficiency gains for data-intensive workloads. The \upmem platform \cite{upmem}, which has been recently acquired by Qualcomm, stands out as the first commercially available PIM architecture, offering a practical, FPGA-free solution that integrates general-purpose RISC cores directly into DRAM modules. Its accessibility and scalability make it an increasingly critical target for both industry prototyping \cite{upmem_dpu_olap,upmem_usecase_UPIS} and academic research \cite{LLJ+23, BKP23, BJS23, BJS23b, KZB+23, CTZ+24, FYF+24, RLG+24, GDG+24, DSI+22, CHC23, DNA+23, LCJ20, CGE+23b, KFF+24}.

\upmem's programming model and software ecosystem, although quite robust, are still evolving toward maturity. Our empirical work reveals that the platform's actual performance often falls short of its theoretical capabilities. Specifically, we have identified the following two core issues that greatly limit \upmem's performance or make it unpredictable: (1)~inefficient compiler-generated code for fundamental operations (e.g., integer multiplication) and (2) lack of NUMA-awareness in PIM allocation, e.g., transferring the data between the host and PIM may require crossing the CPU intra-socket interconnect. Since these inefficiencies are rarely documented, researchers must either tolerate opaque performance characteristics or dedicate substantial effort to low-level programming, diverting attention from algorithmic PIM advancements.

In this paper, we demonstrate that within the \upmem ecosystem there exists a significant performance headroom, which can be exploited through targeted software optimizations. We systematically expose and address critical inefficiencies in the \upmem software stack and play with non-standard programming techniques, achieving the following outcomes:
\begin{enumerate}
\item \emph{Computation optimizations:} By making modifications to the assembly generated by the \upmem compiler, we achieve 1.6--2$\times$ speedup in integer addition and 1.4--5.9$\times$ speedup in integer multiplication, depending on the data type. Our findings reveal that the default implementations of these operations are surprisingly inefficient.

\item \emph{Bit-serial processing:} We demonstrate that bit-serial processing can further improve arithmetic computations on low-resolution data. Our \intf bit-serial dot-product (\bsdp) kernel outperforms the native implementation, which  uses standard \inte addition and multiplication operators, by 2.7$\times$.

\item \emph{Data transfer optimizations:} Through minor, backward-compatible API extensions for DPU allocation, which enable explicit NUMA control and memory channel load balancing, we improve the throughput of host-PIM data transfers by up to 2.9$\times$. These enhancements also dramatically reduce variability in transfer speeds, which is a critical factor for reproducibility in research.
\end{enumerate}


\begin{figure*}[tbh]
    \centering
    \includegraphics[width=0.7\linewidth]{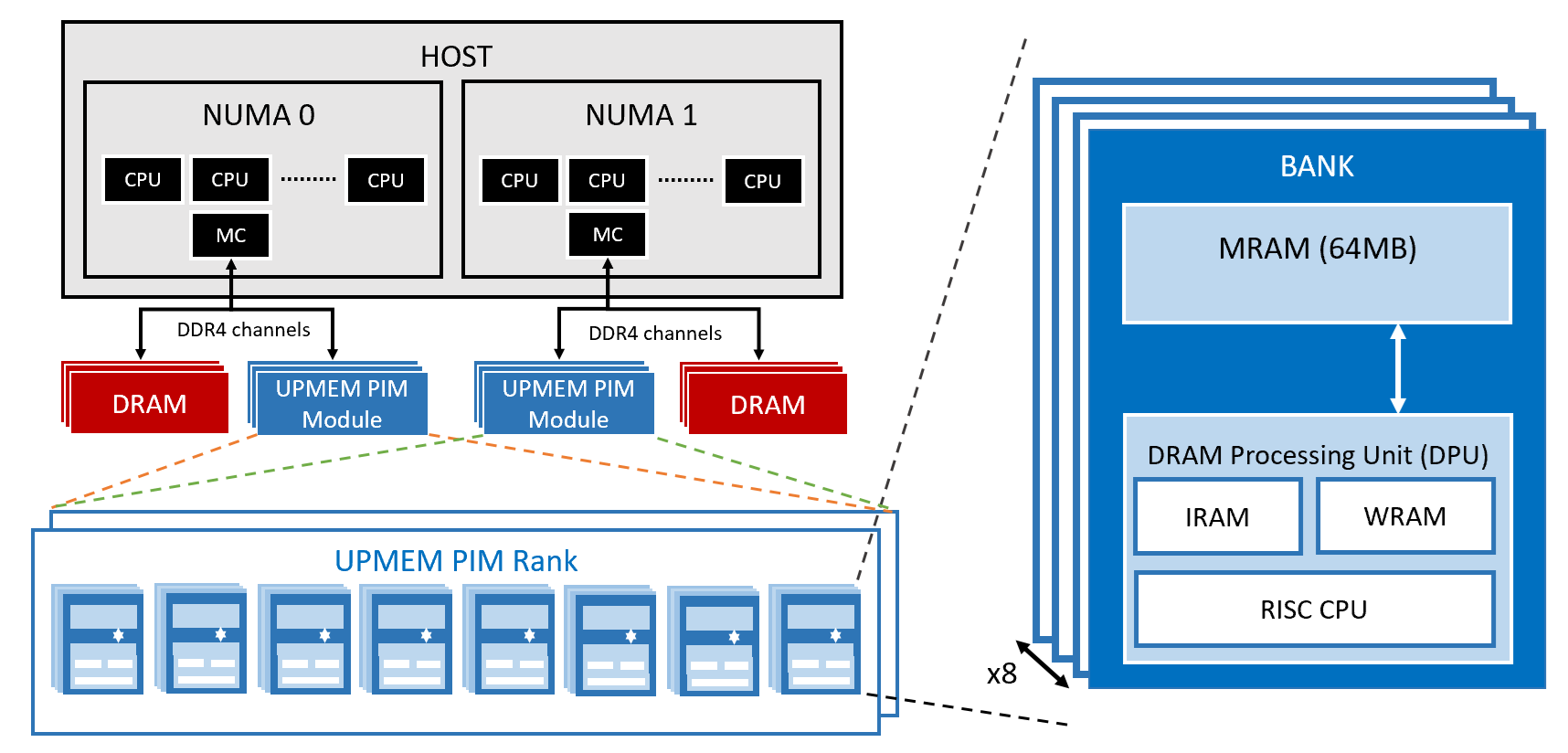}
    \caption{Overview of the \upmem platform architecture}
    \label{fig:upmem-arch}
\end{figure*}

In the paper, we also examine the scalability of \upmem's DPU-based architecture for memory-intensive and bandwidth-bound workloads. We show that when the matrix is preloaded into PIM, a situation common, e.g., in AI model inference, 
our optimized kernels enable UPMEM to outperform a dual-socket CPU server by over 3$\times$ for \inte GEMV and by 10$\times$ for \intf GEMV. Our optimized \inte GEMV kernel outperforms the baseline by 3.5$\times$.

Our work provides a roadmap for unlocking \upmem's latent performance. The optimizations we present require minimal code changes, and are immediately applicable to researchers building upon the platform. By sharing our insights, we aim to transform \upmem from an emerging platform into a robust, predictable, and high-performance solution for PIM research, truly \emph{unleashing} its potential.\footnote{We plan to release the source codes in the near future.}

The rest of this paper is organized as follows. In \Cref{sec:architecture}, we provide an overview of the \upmem's architecture and programming model. We present our arithmetic and data transfer optimizations in Sections \ref{sec:arithmetic_perf}-\ref{sec:mem_transfer}. In \Cref{sec:gemv}, we compare the performance of \upmem with that of a dual‑socket server for GEMV workloads. We review related research in \Cref{sec:related_work} and conclude in \Cref{sec:conclusions}.


\section{\upmem PIM architecture} \label{sec:architecture}

\upmem implements a PIM architecture by integrating general-purpose processing cores directly within DRAM modules at the bank level. An \upmem module\footnote{We focus on the \upmemmodel variant.} is a dual-rank, 8 GB DDR4-2400 DIMM. Each rank comprises 8 PIM chips, collectively providing 64 memory banks per rank. Each 64 MB DRAM bank is coupled with a single \emph{DRAM processing unit (DPU)}, forming the fundamental PIM compute element.

Each DPU is an in-order, 32-bit RISC-based processor core operating at 400~MHz frequency. Its instruction set supports basic integer logical and arithmetic operations. However, it lacks hardware support for more complex operations. Consequently, even \inttt multiplication/division and all floating-point operations must be emulated in software. The core features a 14-stage pipeline and supports up to 16 concurrent hardware threads, also called \emph{tasklets}. The use of tasklets enables programmers to effectively hide memory and pipeline latencies.

Each DPU is connected to its associated DRAM bank, termed \emph{MRAM}, via a private bus. In addition, it has access to a small 24~KB instruction memory (\emph{IRAM}) and a 64~KB SRAM-based scratchpad working memory (\emph{WRAM}). Communication between a host and a DPU is facilitated exclusively through mailbox mechanisms implemented in the MRAM using the DDR protocol. Because DDR commands simultaneously address multiple memory banks,\footnote{In DRAM, a single cache line is distributed across several banks spanning multiple memory chips.} the data must be transposed before it can be transferred between the host (i.e., its CPU caches or DRAM) and MRAM. To this end, the \upmem SDK employs vectorized instructions (x86 AVX \cite{IntelAVX}).

The \upmem SDK provides three modes for data transfer between the host and the DPUs' MRAM:
\begin{itemize}
    \item \emph{Sequential}, which allows one to read from/write to MRAM of a single DPU at a time.
    \item \emph{Parallel}, which is used to  simultaneously access (read or write) the MRAM of multiple DPUs, in order to maximize the utilization of the memory bus.
    \item \emph{Broadcast}, which allows one to push the same data concurrently to the MRAM of multiple DPUs.
\end{itemize}

A typical workflow for employing \upmem is as follows:
\begin{enumerate}
    \item Allocate a specified number of DPUs (or entire ranks of DPUs).
    \item Load the desired kernel onto the allocated DPUs.
    \item Transfer input data from the host to the DPUs.
    \item Execute the kernel on the DPUs.
    \item Transfer the output data back to the host.
\end{enumerate}
Steps (3)–(5) can be executed asynchronously, allowing computation to partially overlap with data transfers. Note that no direct communication between DPUs—even those within the same rank—is possible.

For our experiments, we use an \upmem server featuring a dual-socket Intel Xeon Silver 4216 CPU. Each socket provides six memory channels: one channel connects two standard DDR4-3200 DRAM DIMMs, while the remaining five channels interface with 10 \upmem DIMMs. In total, the system features 256 GB of standard DRAM and 160 GB of PIM, with a cumulative 2560 DPUs available.\footnote{Nine DPUs were found to be faulty and were thus disabled during testing. Hence all our tests are conducted using a total of 2551 DPUs.}

\section{\upmem Computational Performance} \label{sec:arithmetic_perf}


\subsection{Benchmark and baseline results}

We assess the arithmetic performance of a single \upmem DPU using a microbenchmark similar to that in the Processing-In-Memory (PrIM) Benchmark Suite \cite{GEI+22}. This microbenchmark was originally designed to evaluate the performance of addition, subtraction, multiplication, and division operations for \inttt and \float. Given the inherent limitations of the \upmem platform regarding floating-point operations (see \Cref{sec:architecture}), we focus instead on the \inte and \inttt data types. Specifically, we examine the performance of addition and multiplication, as these operations are fundamental to basic kernels, such as GEMV and generalized matrix-matrix multiplication (GEMM).

In the microbenchmark (see \Cref{lst:benchmark}), each tasklet iteratively copies blocks of 1024 values from a 1M-element buffer in MRAM to WRAM, the fast scratchpad memory (line 13). Depending on the selected test mode, the tasklet then performs either scalar addition (line 4) or scalar multiplication (line 6) on each element of the block. Only this computation phase is timed. Performance results are reported in millions of addition or multiplication operations per second (MOPS). To maintain synchronization and ensure uniform progress, tasklets occasionally synchronize on a barrier (lines 14 and 20).





\begin{figure}[t]
\begin{lstlisting}[breaklines=true]
void update(T *buffer, T scalar) {
    for (unsigned int i = 0; 
                i < BLOCK_SIZE / sizeof(T); i++) {
    #ifdef ADD
        buffer[A] += scalar;
    #elif MUL
        buffer[A] *= scalar;
    #endif
    }
}

int main() {
    for (i = 0; i < input_size * sizeof(T); i++) {
        mram_read(mram_base_A + i, wram_A, BLOCK_SIZE);
        barrier_wait(&barrier);
        timer_start(&cycles);

        update(wram_A, scalar);

        result += timer_stop(&cycles);
        barrier_wait(&barrier);
        mram_write(wram_A, mram_base_A + i, BLOCK_SIZE);
    }
}
\end{lstlisting}
\caption{A microbenchmark used to evaluate the arithmetic performance of a single \upmem DPU. Adapted from \cite{GEI+22}. \texttt{BLOCK\_SIZE} is set to 1024.}
\label{lst:benchmark}
\end{figure}

\begin{figure}[tb]
\centerline{\includegraphics[width=\linewidth]{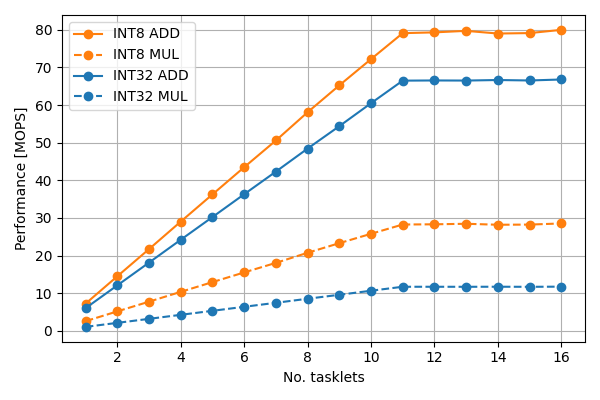}}
\caption{The baseline arithmetic performance of a single DPU. The performance levels off for 11 tasklets, because only 11 out of the 14 pipeline stages can operate concurrently.}
\label{fig:prim_baseline}
\end{figure}

In \Cref{fig:prim_baseline}, we present the baseline results obtained from our microbenchmark. The 6$\times$ lower performance of \inttt multiplication compared to \inttt addition is expected, given the absence of hardware support for \inttt multiplication, and aligns with the findings in \cite{GEI+22}. However, \inte multiplication remains more than 2.7$\times$ slower than \inte addition, despite the presence of a family of dedicated \inte multiplication instructions, such as \mulslsl \cite{upmem2025mulslsl}, in the \upmem ISA. This discrepancy is unexpected, as the specialized multiplication instruction should perform comparably to a similar addition instruction. 
In the following sections, we examine the performance differences between addition and multiplication in greater detail, exploring ways to enhance efficiency. Additionally, we discuss several general programming techniques that can significantly improve computational performance.


\begin{algorithm}[tb]
\caption{Shift-and-add multiplication algorithm}
\label{alg:shift_and_add}
\begin{algorithmic}[1]
\Function{shift\_and\_add\_mul}{a, b}
    \State $acc \gets 0$
    \State $i \gets 0$
    \If{$a < b$} \label{alg:saa:comp-start}
        \State{$swap(a, b)$} \label{alg:saa:comp-end}
    \EndIf
    \While{$b \neq 0$ \textbf{and} $i < 32$}
        \If{$LSB(b) = 1$} \Comment{Least-significant bit is 1} \label{alg:saa:lsb}
            \State $acc \gets acc + (a \ll i)$ \label{alg:saa:acc_add}
        \EndIf
        \State $b \gets b \gg 1$ \Comment{Shift multiplier right} \label{alg:saa:bshift}
        \State $i \gets i + 1$ \label{alg:saa:i_add}
    \EndWhile
    \State \Return $acc$
\EndFunction
\end{algorithmic}
\end{algorithm}


\subsection{Optimizing 8-bit Integer Multiplication} \label{sec:arithmetic_perf:int8}


A closer examination of the microbenchmark's assembly and performance results shows that while \inte addition completes in a single cycle, \inte multiplication does not. Instead of using the native \mulslsl instruction, the compiler generates a call to the \mulsi routine, which implements a shift-and-add algorithm to calculate the product of two \inttt operands. This algorithm decomposes one operand into a sum of powers of two and uses left bit-shifting to perform efficient multiplication of the other operand by the powers of two (see \Cref{alg:shift_and_add}).


The algorithm starts by ensuring that the multiplier ($b$) is the lower value of the two, swapping $a$ and $b$ if necessary (lines \ref{alg:saa:comp-start}-\ref{alg:saa:comp-end}). 
It then iteratively inspects each bit of $b$: for each iteration, it checks the least significant bit (line \ref{alg:saa:lsb}) and, if the bit is set, adds 
$a$ bit-shifted left by the loop counter $i$ (so $a$ multiplied by $2^i$) to the accumulator (line \ref{alg:saa:acc_add}). After examining the bit, the multiplier is right-shifted in preparation for the next loop iteration (line \ref{alg:saa:bshift}). The loop counter is incremented as well (line \ref{alg:saa:i_add}). 

The \mulsi routine does not implement \Cref{alg:shift_and_add} in a straightforward way (we give the decompiled code in \Cref{fig:mulsi3}). Instead, it leverages a special \mulstep instruction \cite{upmem2025mulstep}, which performs all operations from lines \ref{alg:saa:lsb}-\ref{alg:saa:i_add} in a single cycle. More precisely, \mulstep examines the least significant bit of the multiplier, stored in \texttt{d0.low (r0)}, to determine whether to add the left-shifted multiplicand, \texttt{r2}, to the accumulator, \texttt{d0.high (r1)}. The multiplier is then right-shifted in preparation for the next step. Consequently, a full-width \inttt multiplication may require up to 32 \mulstep instructions, while multiplying \inte operands needs at most 9.


    
    

\begin{figure}[tb]
\begin{lstlisting}[language={[RISC-V]Assembler}]
jgtu %2, %1, __mulsi3_swap
move r2, %1
move r0 %2, true, __mulsi3_start
__mulsi3_swap:
move r2, r1
move r0, r0
__mulsi3_start:
move r1, zero
mul_step d0, r2, d0, 0 , z, __mulsi3_exit
mul_step d0, r2, d0, 1 , z, __mulsi3_exit
mul_step d0, r2, d0, 2 , z, __mulsi3_exit
\end{lstlisting}
\vspace{-0.3cm}
...
\vspace{-0.2cm}
\begin{lstlisting}[language={[RISC-V]Assembler}, numbers=left, firstnumber=39]
mul_step d0, r2, d0, 30, z, __mulsi3_exit
mul_step d0, r2, d0, 31, z, __mulsi3_exit
__mulsi3_exit:
move r0, r1
\end{lstlisting}
\caption{The \mulsi routine used on \upmem to perform \inttt multiplication.}
\label{fig:mulsi3}
\end{figure}

\begin{figure}[htbp]
\centerline{\includegraphics[width=\linewidth]{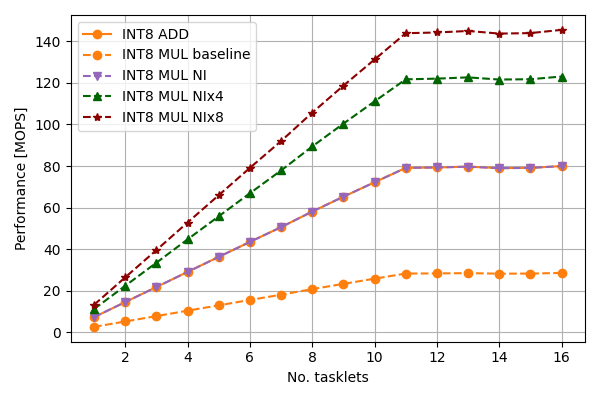}}
\caption{Performance of \inte multiplication on a single DPU, done using the \mulsi routine (baseline), or native instructions (\inteni), and with loading data in 32b (\intenifour) or 64b (\intenieight) blocks. Performance of \inte addition for reference (same as \inte 
\multt \inteni).}
\label{fig:int8_opt}
\end{figure}

Clearly, the use of \mulsi for \inte multiplication is unnecessary, and calls to \mulsi can be replaced with native instructions, such as \mulslsl (which performs multiplication of the lower bytes of values stored in two 32b registers). By ensuring that the correct multiplication instruction is used (denoted \emph{native instruction}, \inteni), we achieve \inte multiplication performance that is on par with \inte addition (see \Cref{fig:int8_opt}).


We can improve the performance of \inte multiplication also using a different technique. Although a tasklet operates on data stored in WRAM, which is several times faster than MRAM, loading data byte-by-byte into DPU registers is inefficient. To this end, we load data in 32-bit (\intenifour) or 64-bit (\intenieight) blocks, as shown in \Cref{lst:loading_data_in_batches}. 
This approach improves \inte multiplication performance by an additional 80\%, achieving almost 5$\times$ speedup compared to the baseline (see \Cref{fig:int8_opt}).

\begin{figure}[htbp]
\begin{lstlisting}[breaklines=true]
    uint64_t d = *((uint64_t*) &bufferA[i]);
    uint32_t l = d;
    uint32_t h = d >> 32;
    int8_t temp;
    __builtin_mul_sl_sl_rrr(temp, l, scalar);
    bufferA[i]     = temp;
    __builtin_mul_sh_sl_rrr(temp, l, scalar);
    bufferA[i + 1] = temp;
    l >>= 16;
    __builtin_mul_sl_sl_rrr(temp, l, scalar);
    bufferA[i + 2] = temp;
    __builtin_mul_sh_sl_rrr(temp, l, scalar);
    bufferA[i + 3] = temp;
    __builtin_mul_sl_sl_rrr(temp, h, scalar);
    bufferA[i + 4] = temp;
    __builtin_mul_sh_sl_rrr(temp, h, scalar);
    bufferA[i + 5] = temp;
    h >>= 16;
    __builtin_mul_sl_sl_rrr(temp, h, scalar);
    bufferA[i + 6] = temp;
    __builtin_mul_sh_sl_rrr(temp, h, scalar);
    bufferA[i + 7] = temp;
\end{lstlisting}
\caption{Loading eight \inte values as a single 64b block and multiplying them by a scalar using the correct \upmem instructions.}
\label{lst:loading_data_in_batches}
\end{figure}

\subsection{Optimizing 32-bit Integer Multiplication}

For \inttt multiplication, the aforementioned \mulsi routine performs up to 32 \mulstep instructions to complete. However, the multiplication can be done more efficiently, once we decompose \inttt operands into byte-size chunks and use the native \uinte multiplication \cite{upmem2025mul} and bit-shift operations with proper sign handling. We refer to this approach as \emph{decomposed \inttt multiplication} (\intttdim).

More precisely, let $X = (x_3, x_2, x_1, x_0)$ and $Y = (y_3, y_2, y_1, y_0)$ represent the byte-level decomposition of the absolute values of the operands. Each $x_i$ and $y_i$ denotes an 8-bit component, with $x_3$ and $x_0$ corresponding to the most and least significant bytes, respectively. 

The sign of the final product $X \cdot Y$ is determined through the \xor operation on the most significant bits:
\[
\text{sign} = \text{msb}(X) \oplus \text{msb}(Y)
\]

\noindent Then: 
\[
X \cdot Y = \begin{cases}
-|X| \cdot |Y| & \text{if } \text{sign} = 1 \\
\ \ |X| \cdot |Y| & \text{otherwise}
\end{cases}
\]

\noindent where $|X| \cdot |Y|$ can be calculated as follows:
\[
\begin{aligned}
X' \times Y' = &\; 2^0 (x_0 \cdot y_0) \\
+ &\; 2^8 (x_0 \cdot y_1 + x_1 \cdot y_0) \\
+ &\; 2^{16} (x_0 \cdot y_2 + x_1 \cdot y_1 + x_2 \cdot y_0) \\
+ &\; 2^{24} (x_0 \cdot y_3 + x_1 \cdot y_2 + x_2 \cdot y_1 + x_3 \cdot y_0)
\end{aligned}
\]

\begin{figure}[tb]
\centerline{\includegraphics[width=\linewidth]{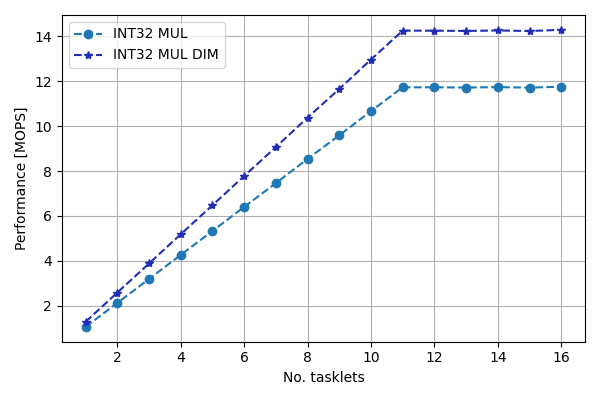}}
\caption{Performance of \inttt multiplication on a single DPU, done using the \mulsi routine (baseline), or through decomposed \inttt multiplication (\intttdim).}
\label{fig:int32_opt}
\end{figure}

The above approach allows us to perform multiplication of a pair of \inttt values in up to 26 cycles, yielding 16\% improvement in our microbenchmark compared to the baseline (see \Cref{fig:int32_opt}).

\subsection{Loop unrolling} \label{sec:arithmetic_perf:loop}

On the \upmem architecture, loop execution introduces significant overhead due to costly jump operations and the lack of branch prediction or speculative execution on DPUs. Consequently, in loops with only a few instructions, control flow overhead can dominate the actual computation time. This is evident in the microbenchmark we used earlier (\Cref{lst:benchmark}): while \inte addition executes in a single cycle, a single loop iteration requires approximately five cycles.

Our experience suggests that the compiler provided with the \upmem SDK is generally conservative in its handling of loop unrolling, due to the limited size of the instruction memory (IRAM) available to each DPU (24~KB). However, sometimes the compiler behaves in a quite unpredictable manner.

We analyzed the assembly generated by the compiler for our microbenchmark (\Cref{lst:benchmark}) while varying the number of loop iterations. For a small number of iterations, the \update function is inlined, the loop is fully unrolled, and the correct \mulslsl instruction is used. As the number of iterations increases to just 29, the function is no longer inlined, though the loop remains fully unrolled, continuing to utilize the efficient one-cycle \mulslsl instruction. However, once the loop has more than 39 iterations, the function is again inlined, the loop is not unrolled, and relies on the costly \mulsi routine. 



\begin{figure}[t]
\centerline{\includegraphics[width=\linewidth]{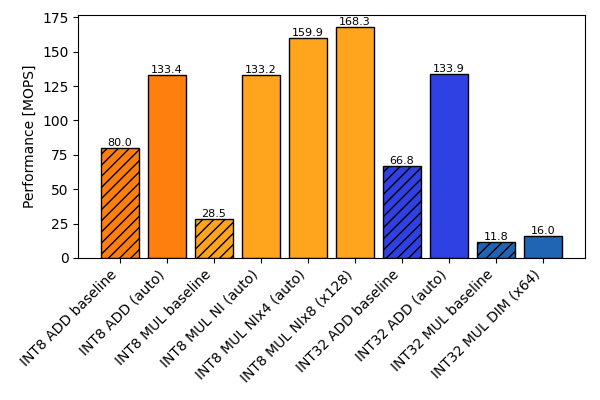}}
\caption{The peak arithmetic (addition and multiplication) performance of a single DPU for the \inte and \inttt data types (orange and blue bars, respectively). The baseline results are given in bars with cross hatch pattern. Optimized variants are due to the use of \texttt{\#pragma unroll} (auto), \texttt{\#pragma unroll 64} (x64) or \texttt{\#pragma unroll 128} (x128).}
\label{fig:unroll}
\end{figure}

We extensively experimented with manual loop unrolling and the use of the \texttt{\#pragma unroll} directive to assess the impact of these techniques on the \upmem performance. \Cref{fig:unroll} illustrates the speedup achieved through loop unrolling across various arithmetic microbenchmarks. The performance gains are especially pronounced for integer addition and simple multiplication operations, underscoring the disproportionate impact of control flow on short loop bodies. Note that applying \texttt{\#pragma unroll} with automatic unrolling or a high unroll count can lead to IRAM overfill, which results in a linker error.

Our results show that \inttt addition benefits the most from loop unrolling, effectively doubling the throughput. Addition and multiplication of \inte (\inte \addtt and \inte \multt \inteni) exhibit similarly strong gains (67\%), confirming that short loops benefit greatly from reduced control overhead. For code with more complex loops, like in our tests of \inte \multt \intenifour and \inte \multt \intenieight, returns diminish sharply (30\% and 16\%).

Unrolling closes the initial performance gap between the performance of \inte and \inttt addition (80 and 67 MOPS, respectively). With our optimizations, in both cases, we can achieve 133 MOPS on a single DPU. However, due to massive improvements in the performance \inte multiplication, the gap with 
the performance of \inttt multiplication increased from the initial 2.4$\times$ to over 10$\times$. The gap remains, even though we have improved the performance of \inttt multiplication by over 35\%.






\begin{algorithm}[t]
\caption{Bit-serial Dot Product of \uintf}
\label{alg:dot_product}
\begin{algorithmic}[1]
\Procedure{DotProduct}{bufferA : array of UINT32,\newline 
\hspace*{3.78cm} bufferB : array of UINT32,\newline
\hspace*{3.78cm} size : UINT32}
    \State $\textit{res} \gets 0$
    \State $\textit{blockIters} \gets \textit{size} / 4$  
    \For{$i \gets 0$ \textbf{to} $\textit{blockIters} - 1$ \textbf{step} 4}
        \Comment{\textit{Unrolled 8$\times$}}
        \State Load $x_{\text{bits}}[0..3]$ from \texttt{bufferA} at indices $i$ to $i+3$ 
        \State Load $y_{\text{bits}}[0..3]$ from \texttt{bufferB} at indices $i$ to $i+3$ 
        \For{$j \gets 0$ \textbf{to} $3$}  \Comment{\textit{Fully unrolled}}
            \For{$k \gets 0$ \textbf{to} $3$}  \Comment{\textit{Fully unrolled}}
                \State $\textit{matches} \gets x_{\text{bits}}[j]\ \andinst\  y_{\text{bits}}[k]$
                \State $\textit{popc} \gets \popcount(\textit{matches})$ \Comment{\texttt{cao}}
                \State $\textit{shift} \gets j + k$  
                \State $\textit{res} \gets \textit{res} + \leftbitshift(\textit{popc}, \textit{shift})$ \label{bsdp:lsl_acc} \Comment{\texttt{lsl\_add}}
            \EndFor
        \EndFor
    \EndFor
    \State \Return $\textit{res}$
\EndProcedure
\end{algorithmic}
\end{algorithm}

\section{Bit-serial dot product}

In this section, we demonstrate how the \upmem platform enables efficient low-precision calculations (e.g., \uintf or \intf) using bit-serial processing. Specifically, we showcase the bit-serial computation of a dot product (\bsdp), as originally proposed in \cite{UJ17, MKS23}. For clarity, we adopt \uintf as the base data type. Later we discuss how the algorithm can be extended to accommodate signed integers (\intf).

\subsection{Principles of bit-serial dot product}

Let $A$ and $B$ be two \uintf vectors, whose dot product $A \cdot B$ we want to calculate. In bit-serial processing, each element of $A$ and $B$ is treated as a sum of its bits scaled by powers of 2:
\begin{equation}
A[i] = \sum_{j=0}^{3} a_j^{(i)} \cdot 2^j, \quad 
B[i] = \sum_{k=0}^{3} b_k^{(i)} \cdot 2^k
\end{equation}

\noindent Then the dot product $A \cdot B$ can be calculated as follows:
\begin{equation}
A \cdot B = \sum_{i=0}^{N-1} \left( A[i] \cdot B[i] \right) 
= \sum_{i=0}^{N-1} \left( 
  \underbrace{
    \sum_{j=0}^{3} \sum_{k=0}^{3} 
    \left( a_j^{(i)} \cdot b_k^{(i)} \cdot 2^{j+k} \right)
  }_{\text{Bit-level products}}
\right)
\end{equation}

\noindent We can reorganize the summation order to facilitate parallel computation:

\begin{equation}
A \cdot B = \sum_{j=0}^{3} \sum_{k=0}^{3} \underbrace{2^{j+k}  \cdot}_{\text{(1)}} 
  \left( \underbrace{\sum_{i=0}^{N-1} a_j^{(i)} \cdot b_k^{(i)}}_{\text{(2)}}\right)
\end{equation}

\subsection{Bit-serial dot product on \upmem}

Now, we can make the following two key observations regarding \upmem's ISA:  
\begin{itemize}
    \item The multiplication by $2^{j+k}$ (1) can be efficiently performed using a left bit shift (\leftbitshift) operation, eliminating the need for explicit multiplication instructions.  
    \item The summation term (2) independently evaluates all bit-planes $(i)$ of $A$ and $B$. Therefore, if each bit-plane $(i)$ of $A$ and $B$ were stored as a separate bit vector, we could compute the term by performing a bitwise \andinst operation and then counting the number of ones in the resulting bit vector. This operation, commonly known as \emph{population count} (\popcount), can be efficiently executed on \upmem using the \texttt{cao} instruction, which computes the population count of a 32-bit register \cite{upmem2025cao}.  
\end{itemize}

\begin{figure}[tbp]
\center
\includegraphics[scale=0.50]{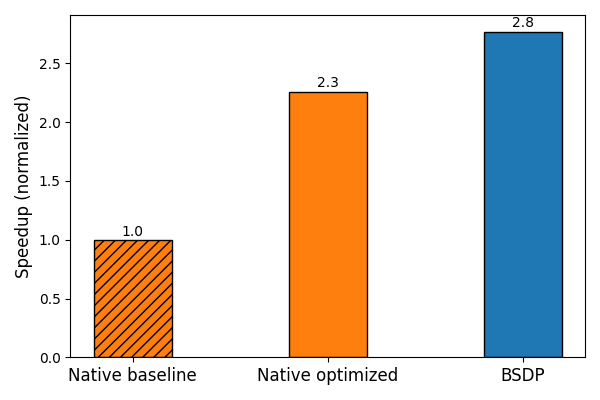}
\caption{Performance of bit-serial dot product (\emph{\bsdp}) of \intf values on a single DPU normalized to the native dot product performance achieved with each \intf stored as a separate \inte and using \uinte \texttt{\addinst} and \texttt{\mulinst} instructions (\emph{native baseline}). An optimized version of the baseline kernel (\emph{native optimized}) features improvements discussed in \Cref{sec:arithmetic_perf:int8} and \Cref{sec:arithmetic_perf:loop}.}
\label{fig:bsdp_results}
\end{figure}


To enable efficient \bsdp on \upmem, we first transpose the $A$ and $B$ vectors as follows: every block of 32 elements in $A$ (and similarly in $B$) is stored as four consecutive \uinttt values within a buffer. The first value represents the $2^0$ bit-plane of the values from the block, the second value corresponds to the $2^1$ bit-plane, and so on. This structured arrangement greatly simplifies the computation of the dot product $A \cdot B$, as illustrated in \Cref{alg:dot_product}. 

For improved performance, we make extensive use of loop unrolling and optimize memory accesses by loading data in 64-bit blocks (see also \Cref{sec:arithmetic_perf}). Additionally, on \upmem, the operations in \cref{bsdp:lsl_acc} can be efficiently executed using a single instruction, \texttt{lsl\_add}, which eliminates unnecessary overhead and enhances computational efficiency \cite{upmem2025lsladd}.

To support signed integers, if exactly one of $j$ or $k$ equals 3, then on line 12 the accumulator \emph{res} should be decremented, rather than incremented, by  $\leftbitshift(\textit{popc}, \textit{shift})$ \cite{UJ17}. Since we use full loop unrolling, there are not additional checks, and the performance is the same for both \intf and \inte cases.

We expect \bsdp to be used on \upmem to repeatedly perform GEMV of the same matrix $M$ and different vectors $x$ (in fact, we evaluate such case in \Cref{sec:gemv:int4}). It means that matrix $M$ can be kept in \upmem's MRAM across multiple invocations of GEMV. Since the transposition of $M$ to accommodate bit-serial processing only occurs once, its cost is amortized over multiple GEMV executions. The transformation of data layout can be efficiently performed on the host using vector instructions, such as AVX512. The cost of transposing vector $x$ is negligible compared to the overhead of broadcasting it to the DPUs. The above argument extends also to repeated GEMM invocations.

\subsection{Performance results}

In \Cref{fig:bsdp_results}, we present the performance of our \bsdp kernel on \intf. We achieve a 2.7$\times$ speedup over a baseline implementation that stores each \intf value as a single \inte and computes the dot product with native \inte \texttt{\addinst} and \texttt{\mulinst} instructions.\footnote{Storing two \intf values per byte requires costly unpacking operations, which noticeably slow down the dot-product calculation.} Even when the \inte-arithmetic and loop-level optimizations from \Cref{sec:arithmetic_perf:int8} and \Cref{sec:arithmetic_perf:loop} are applied to the baseline kernel, our \bsdp implementation still maintains a 22\% performance advantage.

\section{Data transfer optimizations}
\label{sec:mem_transfer}

\subsection{Bottlenecks in host-PIM data transfers}

On the \upmem platform, host-PIM data transfers can incur noticeable overhead due to several limiting factors. Firstly, the \upmem modules are based on DDR4-2400, which means that the theoretical maximum transfer speed is capped at 19.2 GB/s per memory channel. In our server, each memory channel designated for \upmem supports a pair of \upmem DIMMs; a total of 20 \upmem DIMMs are distributed across two CPU sockets (see also \Cref{sec:architecture}). Secondly, since \upmem relies on the standard DDR protocol, which by design operates across multiple memory banks simultaneously, the layout of the data must be changed before the data can be transferred between the host and the \upmem modules. The change of the data layout, which is performed with vectorized instructions, places a heavy load on the CPU, especially when the data is transferred to multiple \upmem DIMMs concurrently. Thirdly, for each CPU socket, there is only a pair of DRAM DIMMs, which are connected to a single memory channel. Despite DRAM operating at DDR-3200 speeds, which are significantly faster than DDR4-2400, it is often the DRAM that becomes the bottleneck in host-\upmem data transfers. 

The performance limiting factors mentioned above stem directly from the hardware design and server configuration, making efficient resource utilization critical. Our experiments show that inefficient DPU allocation further slows data-transfer performance. 


In the current version of the \upmem SDK (2025.1.0), the DPU allocation is implemented in a straightforward manner. Firstly, the \upmem SDK retrieves a complete list of DPUs via the \texttt{libudev} API \cite{libudev}.\footnote{The order of devices in the list remains consistent across machine restarts.} The DPUs are grouped into ranks, mirroring the configuration on the \upmem DIMMs. The driver then iterates over ranks and the available DPUs within each rank, and allocates the requested number of DPUs.

It is important to note that this allocation process is not NUMA-aware: the allocated DPUs may reside on \upmem DIMMs attached to either (or both) CPU sockets. Consequently, data transfer speeds between the host and \upmem modules vary greatly across benchmark runs. The performance is noticeably better when all allocated DPUs are located on the same CPU socket as the DRAM DIMM designated for data storage (we quantify this difference below). Currently, \texttt{numactl} \cite{numactl} cannot be used to constrain DPU allocation to a specific CPU socket.

The DPU allocation process also abstracts away the physical location of DPUs relative to the memory channel used for communication. This means that, for example, when 256 DPUs (four full \upmem ranks) are allocated, they could end up on a pair of \upmem DIMMs connected via the same memory channel to the CPU, thus severely limiting data transfer speeds between the host and the corresponding MRAM. In contrast, the allocation of DPUs from four separate \upmem modules, each connected to the CPU through a different memory channel, allows a high level of parallelism in data transfers. 

\subsection{\upmem SDK extensions}

We extended the \upmem SDK to support NUMA- and rank-location-aware DPU allocation. A sample code demonstrating the extended API is provided in \Cref{lst:numaaware_dpu_allocation}. The new function \texttt{alloc\_buffer\_on\_cpu} enables DRAM buffer allocation on a NUMA node specified (NUMA node 0 in lines 10 and 15, and NUMA node 1 in lines 11 and 16). Additionally, we modified the \texttt{dpu\_alloc\_ranks} function (lines 25-26) to allow the programmer to restrict DPU allocation to only those \upmem ranks, which are connected through specified memory channels to a particular CPU (NUMA node). We balance the allocation of DPUs across all available memory channels (lines 21-22). Our modifications were confined exclusively to the userspace \upmem library and required only 15 additional lines of code.

\begin{figure}[tb]
\begin{lstlisting}
// An example number of UPMEM ranks (64 DPUs each)
// to allocate
uint32_t ranks = 4;

// Handles for DPUs assigned to each NUMA node
dpu_set_t set[2]; 

// Allocate input buffers on each NUMA node
uint64_t *input_buffer[2];
input_buffer[0] = alloc_buffer_on_cpu(0);
input_buffer[1] = alloc_buffer_on_cpu(1);

// Allocate output buffers on each NUMA node
uint64_t *output_buffer[2];
output_buffer[0] = alloc_buffer_on_cpu(0);
output_buffer[1] = alloc_buffer_on_cpu(1);

// Make sure to balance the allocation of DPUs 
// across all available memory channels
int32_t *ch[2];
ch[0] = equal_channel_distribution(ranks/2, 0);
ch[1] = equal_channel_distribution(ranks/2, 1);

// Allocate the DPUs
dpu_alloc_ranks(ranks/2, NULL, set, 0, ch[0]);
dpu_alloc_ranks(ranks/2, NULL, set, 1, ch[1]);

// Transfer input data from the host to UPMEM
transfer_data(input_buffer[0], set[0]);
transfer_data(input_buffer[1], set[1]);

// Launch the kernel on the allocated DPUs
dpu_launch(set[0], DPU_SYNCHRONOUS);
dpu_launch(set[1], DPU_SYNCHRONOUS);

// Transfer the output data from UPMEM to the host
transfer_output_data(output_buffer[0], set[0]);
transfer_output_data(output_buffer[1], set[1]);
\end{lstlisting}
\caption{NUMA- and rank-location-aware DPU allocation.}
\label{lst:numaaware_dpu_allocation}
\end{figure}

\subsection{Performance results}

We assess the impact of NUMA and rank-location-aware DPU allocation on data transfer speeds using a simple microbenchmark. In each run, the microbenchmark copies, in the parallel mode, a buffer from DRAM to \upmem (\emph{host-to-PIM}) or vice versa (\emph{PIM-to-host}). Data is transferred in large, 32~MB blocks, for optimal performance \cite{GEI+22}. The number of allocated \upmem ranks varies from 2 to 40, spanning from a single \upmem DIMM to 20 \upmem DIMMs.

\begin{figure}[tb]
\vspace{-0.7cm}
\centerline{\includegraphics[width=\linewidth]{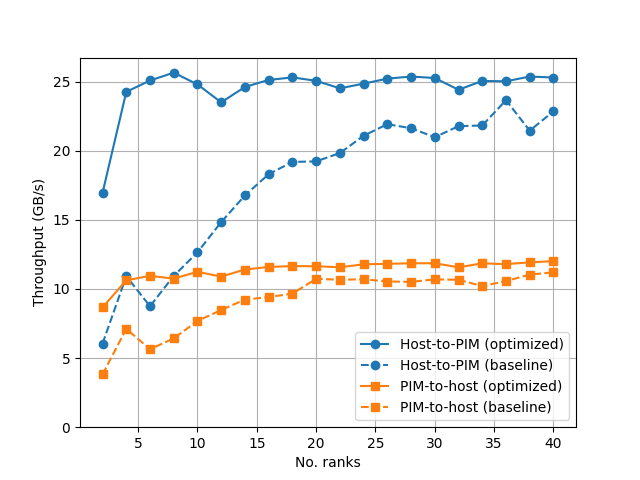}}
\caption{Throughput of data transfer, in the parallel mode, between the host and \upmem as a function of the number of \upmem ranks allocated. }
\label{fig:data_copy}
\end{figure}

We present the microbenchmark results in \Cref{fig:data_copy}. Let us first focus on the stark difference in the performance of the host-to-PIM (blue lines) vs PIM-to-host (orange lines) data transfers, for both our approach (solid lines) and the baseline (dashed lines). The difference stems from the use of fast asynchronous write and slow synchronous read x86 AVX instructions \cite{IntelAVX} to change the layout of the data during the data transfer operations to and from \upmem, respectively.

Note that for both host-to-PIM and PIM-to-host transfers, our NUMA-aware approach with memory channel balancing (solid lines) performs significantly better and achieves peak data transfer throughput with just four allocated \upmem ranks. These \upmem ranks reside on four separate \upmem DIMMs, each connected via separate memory channels, with two channels served by one CPU socket and two by the other. This way we maximize parallel data transfer over the memory bus. Increasing throughput further, by allocating additional \upmem ranks and memory channels, is not possible due to the computational overhead required to reformat data between the DRAM-native and \upmem-native layouts.

The benefits of our approach are most pronounced with a small number of ranks (2–10), with improvements reaching up to 2.9$\times$ (2.4$\times$ on average) for host-to-PIM transfers and 2.3$\times$ (1.8$\times$ on average) for PIM-to-host transfers. In contrast, under the baseline DPU allocation method, even on an otherwise idle system, all allocated ranks reside on only 1–3 \upmem DIMMs attached to the same NUMA node.

The performance benefit of our DPU allocation method decreases with the number of DPUs allocated, falling to roughly 15\% for the host-to-PIM transfers and 10\% for the PIM-to-host transfers. Notably, our method delivers significantly more stable data transfer speeds: repeated tests show variations as little as 0.3~GB/s, compared to fluctuations of 2–4~GB/s observed with the baseline.

\section{Upmem in memory-bound workloads} \label{sec:gemv}

In previous sections, we discussed several optimizations that enhance \upmem's performance for basic arithmetic operations and data transfers. Building on that foundation, in this section, we investigate the performance achievable by \upmem when handling more complex, memory-bound workloads. Our aim is to assess how far \upmem's DPU-based architecture can scale for intensive tasks and to understand the relative impact of data transfer overhead in practical scenarios. Moreover, to contextualize the performance levels achieved by \upmem, we compare its results against those of a modern server.\footnote{A comparison with contemporary GPUs or NPUs is not meaningful, as \upmem remains a prototype platform.}

\subsection{Benchmarks}

We implemented optimized \inte and \intf generalized matrix-vector multiplication (GEMV) kernels \cite{BPP+02} for the \upmem platform. The input matrix is automatically partitioned across all 2551 DPUs by assigning each DPU a contiguous block of rows, while the input vector is broadcast to every DPU. This scheme maximizes data‐parallel throughput with minimal host‐DPU coordination.

For the \intf variant, the host pre-encodes both the matrix and vector into BSDP-compatible formats before launching the kernel. We exclude encoding time from our performance results because (1) matrix encoding is amortized over many inference requests in realistic AI pipelines and (2) the per-vector encoding overhead is negligible compared to UPMEM’s kernel-launch latency.

Furthermore, our experimental setup considers two distinct scenarios for the \upmem platform. In the first case, \emph{GEMV-MV}, we measure the total execution time that includes the data transfer of both the \emph{matrix} and the \emph{vector} inputs, the kernel computation, and the transfer of results back to the host. This comprehensive timing reflects workloads where data movement forms a significant part of the execution profile. In the second case, \emph{GEMV-V}, we assume that the matrix is already resident in \upmem memory (a situation common, e.g., in AI model inference) and focus solely on the transfer of the \emph{vector}, the computation time, and retrieval of the output. This scenario isolates the kernel execution performance and highlights the benefits of a preloaded memory state.

By running tests with matrices and vectors of various sizes, we systematically evaluate the relationship between input size and data transfer overhead on \upmem. This approach provides deeper insights into the intrinsic trade-offs between computation and communication in this novel hardware architecture. For \inte GEMV, we run similar computation on our server using the Arm Compute Library \cite{arm_compute_library}.\footnote{Arm Compute Library does not support \inttt GEMV. This is why we do not include tests for this scenario in our paper.} For \intf GEMV, we adopted Arm-NEON enabled code from llama.cpp \cite{llamacpp}. The server is configured with a dual-socket, Huawei Kunpeng 920 CPU (128 cores in total) and 1024~GB of DDR4-3200 DRAM. 



\begin{figure*}[tb]
  \centering
  \begin{subfigure}[t]{0.49\linewidth}
    \centering
    \includegraphics[width=\linewidth]{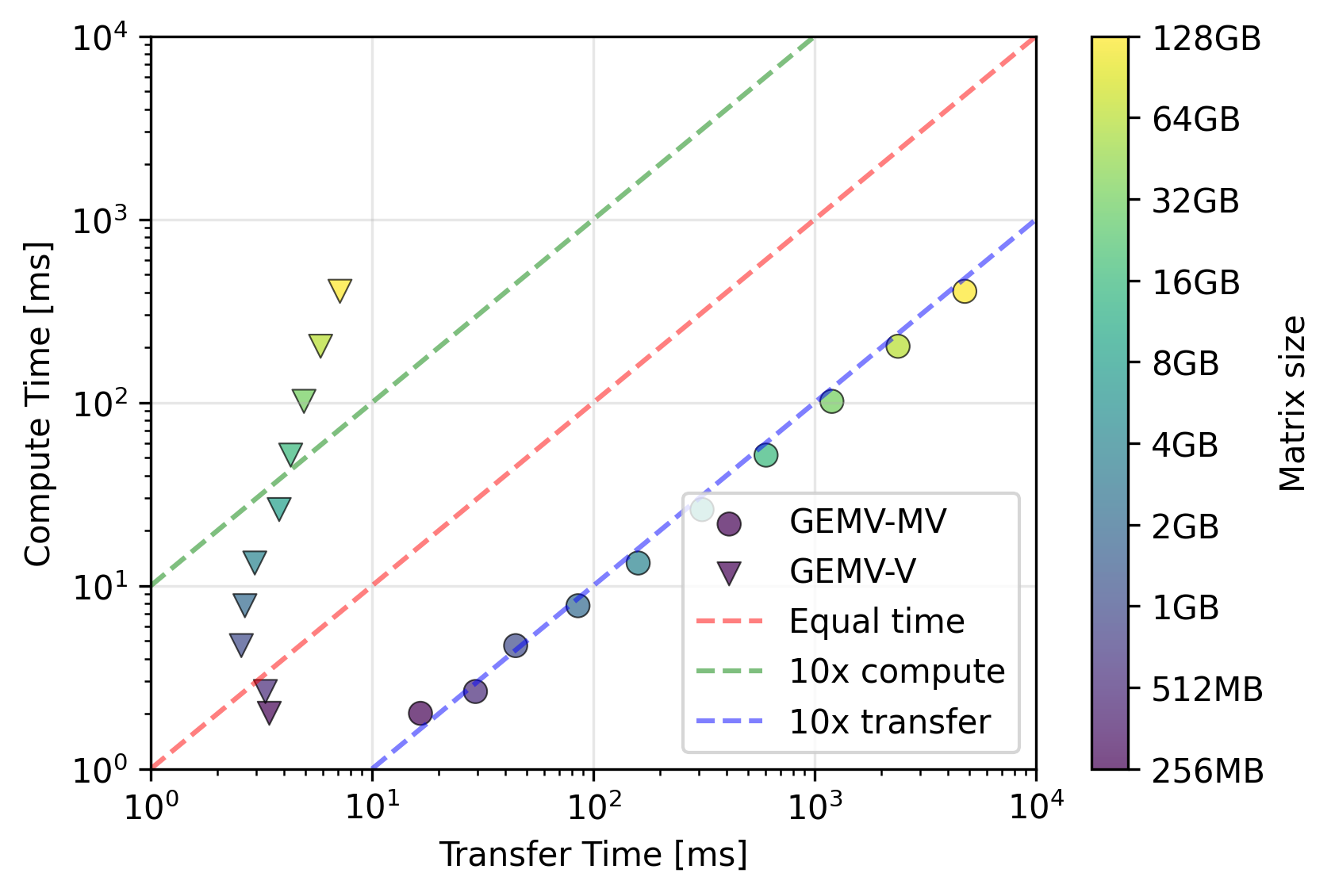}
    \caption{\inte GEMV}
    \label{fig:int8_transfer_compute}
  \end{subfigure}
  \hfill
  \begin{subfigure}[t]{0.49\linewidth}
    \centering
    \includegraphics[width=\linewidth]{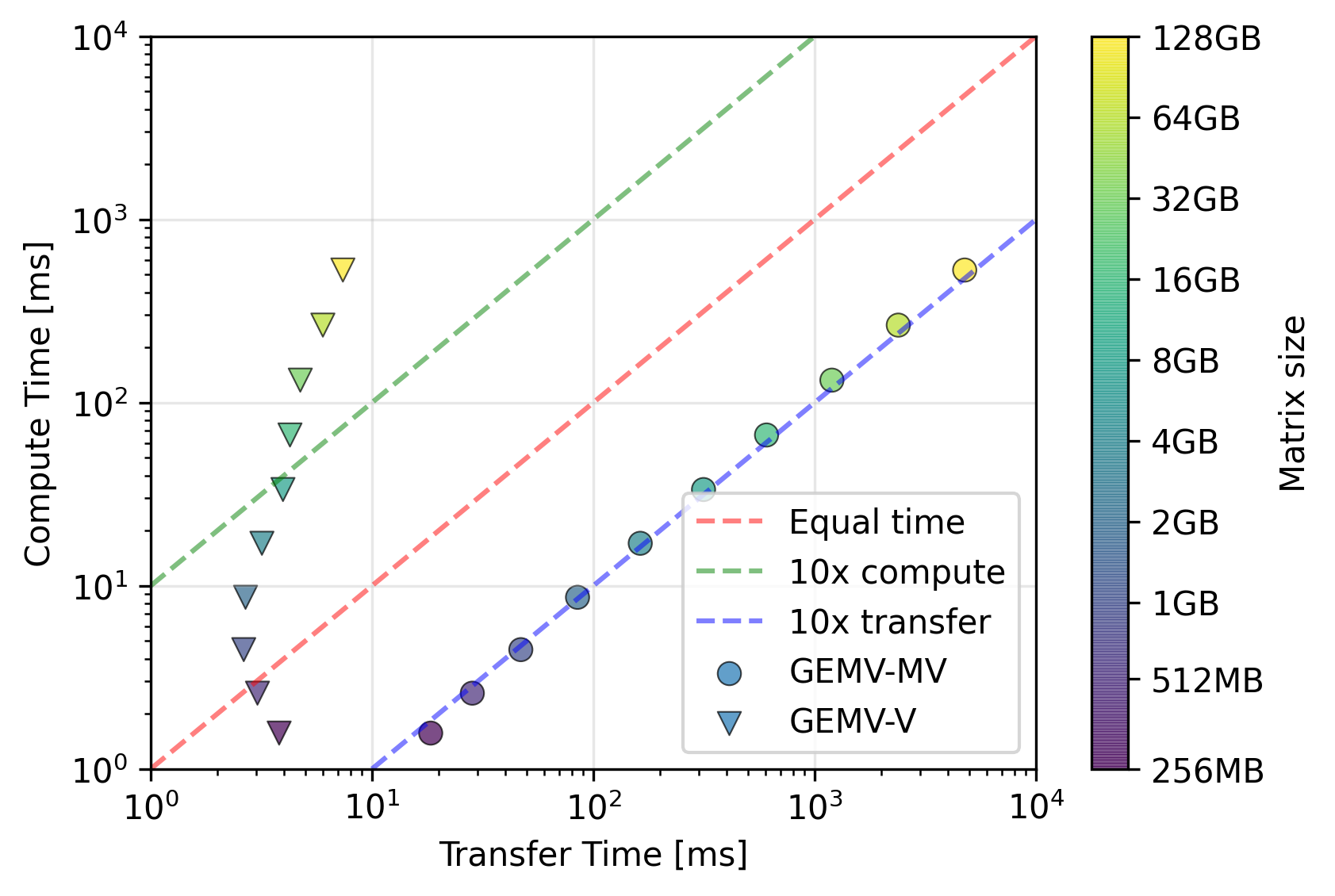}
    \caption{\intf GEMV}
    \label{fig:int4_transfer_compute}
  \end{subfigure}
  \caption{GEMV compute time vs data transfer time on 2551 DPUs. Matrix size ranges from 256~MB to 128~GB.}
  \label{fig:comp_vs_transfer}
\end{figure*}




\begin{figure*}[tb]
  \centering
  \begin{subfigure}[t]{0.49\linewidth}
    \centering
    \includegraphics[width=\linewidth]{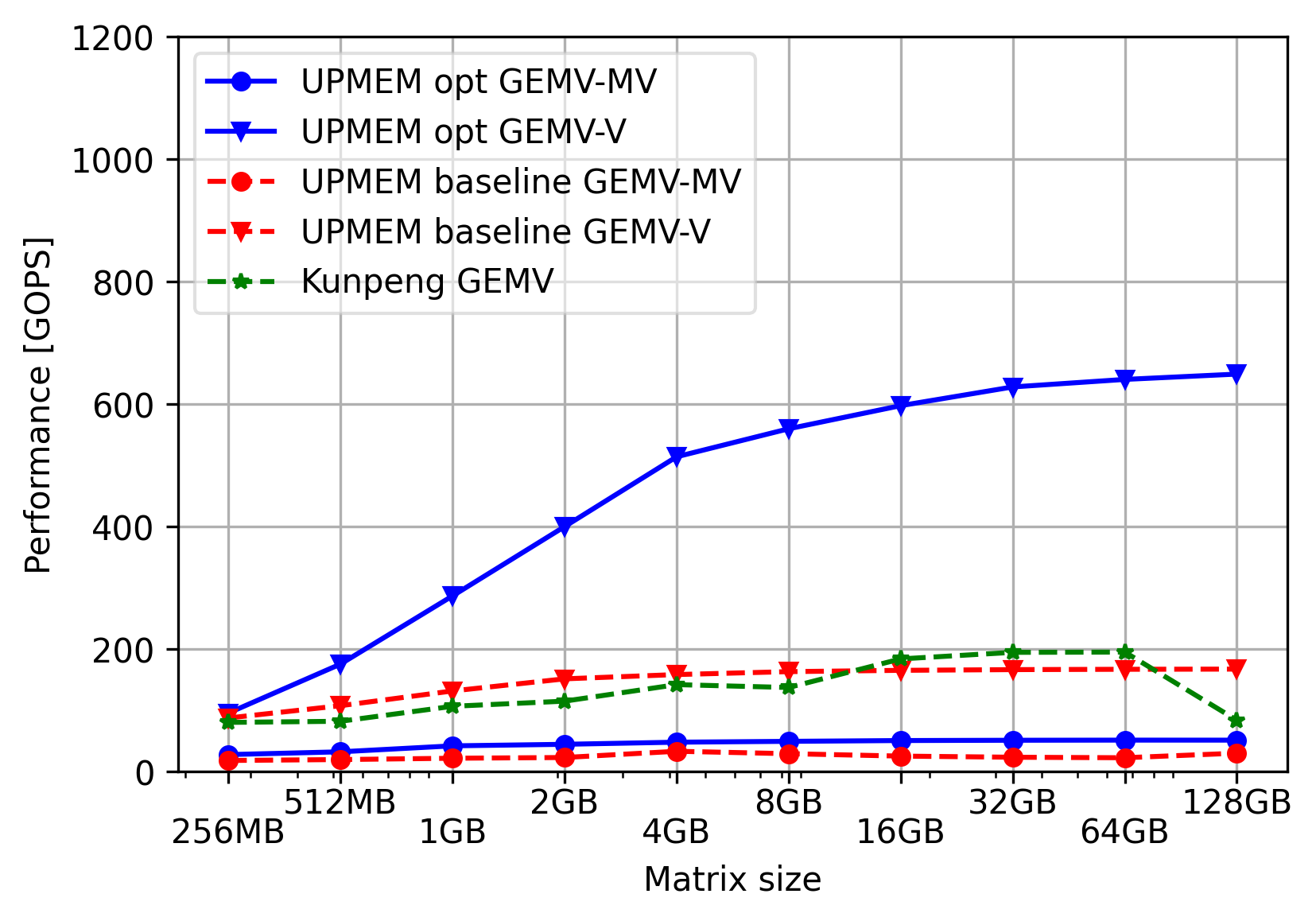}
    \caption{\inte GEMV}
    \label{fig:gemv_int8_perf}
  \end{subfigure}
  \hfill
  \begin{subfigure}[t]{0.49\linewidth}
    \centering
    \includegraphics[width=\linewidth]{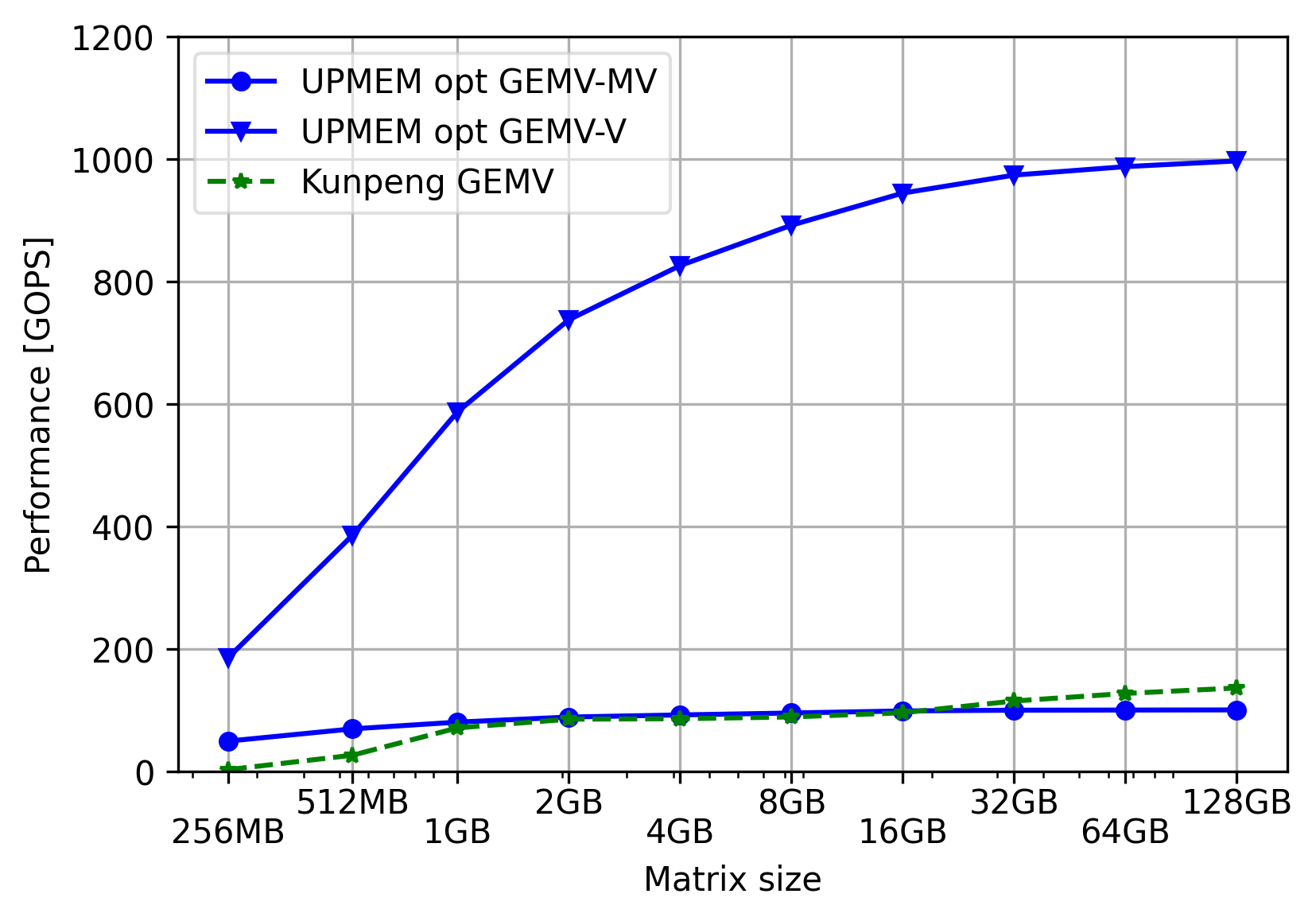}
    \caption{\intf GEMV}
    \label{fig:gemv_int4_perf}
  \end{subfigure}
  \caption{GEMV performance on \upmem (2551 DPUs) and a dual socket Huawei Kunpeng server.}
  \label{fig:gemv_inte_perf}
\end{figure*}


\subsection{\inte GEMV performance}

\Cref{fig:int8_transfer_compute} compares the relative performance of \upmem under two scenarios: GEMV-MV (circles) and GEMV-V (triangles). In the GEMV-MV scenario, the overall execution is dominated by data transfer overhead: the data transfer takes roughly ten times longer than the computation itself. Notably, this behavior persists regardless of the input matrix size, which ranges from 256~MB to 128~GB. These results highlight that efficient data reuse is critical for attaining reasonable performance levels on \upmem.

In contrast, the GEMV-V scenario, where the matrix is preloaded into \upmem's memory banks, shows that computation time significantly exceeds the time spent on transferring input and output data. For a 128~GB matrix, the computation takes approximately 0.4 seconds, which is 57$\times$ more than the transfer time. In practice, transferring the input or output vectors takes only about 2–7~ms, so amount of time that becomes negligible once each DPU processes more than a few megabytes of data (i.e., when the matrix size exceeds around 8~GB). This short transfer time can be regarded as a fixed overhead associated with launching a kernel on \upmem.


We also compared \upmem's performance with that of our Kunpeng server (see \Cref{fig:gemv_int8_perf}). The server tops out at about 200~GOPS. We observed that the GEMV performance using the Arm Compute Library is highly sensitive to matrix dimensions.\footnote{We chose the matrix dimensions so that we could easily tile the matrix across all the DPUs on \upmem.} This sensitivity explains the drop in performance for the 128~GB matrix. However, in all tests, the server's performance never exceeded 220~GOPS.


In the GEMV-MV scenario, the server outperforms \upmem by roughly 4$\times$ for the optimized kernel and about 6$\times$ for the baseline kernel. These results illustrate the significant penalty incurred when the matrix must be transferred every time a kernel is launched on \upmem.
 
However, when the matrix is preloaded (GEMV-V), with our optimizations \upmem achieves higher performance than the server across all tested matrix sizes. \upmem scales all the way up to 650~GOPS, which is over 3$\times$ the peak performance of our server. Without the optimizations, \upmem barely matches the performance of the server in this scenario. These results highlight the great potential of the \upmem platform, unlocked through extensive optimizations.

\subsection{\intf GEMV} \label{sec:gemv:int4}

For \intf GEMV, the compute-to-transfer ratio closely mirrors its \inte counterpart (see \Cref{fig:int4_transfer_compute}). However, in terms of raw performance, our optimized \upmem kernels built on BSDP  deliver marked gains over \inte GEMV (see \Cref{fig:gemv_int4_perf}). GEMV-V peaks at 1000 GOPS, i.e., 1.53$\times$ the throughput of \inte GEMV-V, while GEMV-MV tops out at about 100 GOPS, doubling \inte GEMV-MV performance and closely matching our Kunpeng server. We omit an \upmem baseline for \intf GEMV since the platform lacks direct support for \intf. The Kunpeng server’s \intf performance is about half its \inte throughput and about 10$\times$ lower compared to \upmem opt GEMV-V, reflecting the CPU's absence of native support for the \intf arithmetic and the overhead of packing values in pairs within each byte.

\section{Related work} \label{sec:related_work}

G\'omez-Luna \etal \cite{GEI+22} conducted the first comprehensive analysis of the \upmem platform's performance. Their work introduced a set of microbenchmarks along with \emph{PrIM}, a benchmark suite designed for PIM platforms, to evaluate workloads across various domains. The microbenchmarks assessed the performance of fundamental arithmetic operations--—addition, subtraction, multiplication, and division—--for multiple data types, including \inttt, \intsf, \float, and \double, on a single \upmem DPU. The study revealed that while addition and subtraction are executed efficiently, multiplication and division suffer significantly due to the absence of native hardware support (see also \Cref{sec:architecture}). Recognizing this limitation, the authors advocated for optimized software library routines to emulate these complex operations more effectively. In this work, we specifically investigate the performance of addition and multiplication for the \inttt and \inte data types, which are particularly critical in the performance of quantized AI models. 

In the same paper, the authors claim that \inttt GEMV computation is several dozen times faster on \upmem than on a CPU. However, their evaluation relies on a basic, OpenMP-based GEMV implementation and benchmarks a relatively weak, four-core Intel Xeon processor. In contrast, we assess the \inte GEMV performance on \upmem and compare it against a modern ARM-based server running a state-of-the-art GEMV kernel implementation \cite{arm_compute_library}.

G\'omez-Luna \etal \cite{GEI+22} also analyzed host-\upmem data transfer performance, though primarily in the context of a single \upmem rank. The overhead associated with data transfers on the \upmem platform has also been examined in \cite{LHKR24, FLS23}. Specifically, the authors of \cite{LHKR24} highlight synchronization challenges when memory is accessed by both the CPU and DPUs, as well as the incompatibility of existing memory interleaving techniques with PIM processing. In addition, they evaluate the computational overhead of data layout transformations, which are performed during transfers. A discussion on the full-system perspective on the performance of the \upmem platform, including data transfer and PIM reconfiguration cost, can be found in \cite{FLS23}. The authors also extend the PrIM benchmark suite to account for these factors. Based on the above insights, our work focuses on data transfer performance across the entire server, which features a two-socket NUMA architecture along with multiple DRAM and \upmem DIMMs. 


Kim \etal \cite{KKP+25} introduce \emph{PIM-LLM}, an end-to-end framework designed to accelerate LLMs on PIM platforms through an efficient tiled GEMM library.\footnote{The source code for PIM-LLM appears not to be publicly available.} Since their experiments are conducted on \upmem, they primarily focus on models quantized using \inte. The authors also explore several low-level optimizations specific to \upmem. Notably, they propose memory layout adjustments to minimize host-\upmem transfers and enhance parallel transfer efficiency. Additionally, they introduce PIM resource pooling to reduce the overhead associated with DPU allocation and deallocation. %

Bit-serial processing has emerged as a promising technique for optimizing low-precision arithmetic in constrained environments. Notably, \cite{UJ17, MKS23} demonstrate how 
to calculate dot products with only basic operations, such as \texttt{ADD}, \texttt{AND}, \texttt{BIT-SHIFT}, and \texttt{POPCOUNT}.
These techniques allow one to speed up low-precision GEMV and GEMM calculation, particularly when hardware provides limited support for arithmetic operations (as in case of \upmem). These methods were primarily validated on FPGAs or custom accelerators, and their applicability to emerging PIM architectures, such as \upmem, remained so far underexplored.

Building on these techniques, our work presents further optimizations that enhance the performance of fundamental kernels—such as GEMM—utilized in PIM-LLM 
and other computation-intensive applications on \upmem, which target, among others, analytics and database workloads \cite{LLJ+23, BKP23, BJS23, BJS23b, KZB+23}, machine learning \cite{CTZ+24, FYF+24, RLG+24, GDG+24, DSI+22}, and bioinformatics \cite{CHC23, DNA+23, LCJ20}.


\section{Conclusions} \label{sec:conclusions}

In this paper, we have demonstrated that the \upmem platform, despite its impressive theoretical capabilities, suffers from critical inefficiencies in both computation and data transfer that can hinder its real-world performance. By systematically analyzing \upmem performance and gradually improving the code, we achieved a significant speedup in integer addition and multiplication (1.6--2$\times$ and 1.4--5.9$\times$, respectively)---core operations that previously imposed a surprising computational overhead. We also demonstrated that bit-serial processing of low precision data is a viable option for \upmem: in \inte dot-product calculation, we achieved over 2.7$\times$ speedup over the baseline. Alongside this, our introduction of minor API enhancements to enforce NUMA awareness improved host-PIM data transfer throughput by up to 2.9$\times$, while substantially reducing performance variability. We also showed that when the matrix is preloaded into PIM, our optimized kernels enable UPMEM to outperform a dual-socket CPU server by over 3$\times$ for \inte GEMV and by 10$\times$ for \intf GEMV. Our optimized \inte GEMV kernel outperforms the baseline by 3.5$\times$.

Collectively, our work transforms \upmem into a more robust and predictable platform for PIM research and provides a clear roadmap for further enhancements. Future efforts could extend these techniques to other operations and more advanced memory management strategies.

\bibliographystyle{IEEEtranS}
\bibliography{ccmcc2025}

\end{document}